\documentstyle[12pt]{article}
\begin{document}
\thispagestyle{empty}

\newcommand{\be}{\begin{equation}}
\newcommand{\ee}{\end{equation}}
\newcommand{\sect}[1]{\setcounter{equation}{0}\section{#1}}
\renewcommand{\theequation}{\thesection.\arabic{equation}}
\newcommand{\vs}[1]{\rule[- #1 mm]{0mm}{#1 mm}}
\newcommand{\hs}[1]{\hspace{#1mm}}
\newcommand{\mb}[1]{\hs{5}\mbox{#1}\hs{5}}
\newcommand{\bea}{\begin{eqnarray}}
\newcommand{\ena}{\end{eqnarray}}

\newcommand{\wt}[1]{\widetilde{#1}}
\newcommand{\und}[1]{\underline{#1}}
\newcommand{\ov}[1]{\overline{#1}}
\newcommand{\sm}[2]{\frac{\mbox{\footnotesize #1}\vs{-2}}
		   {\vs{-2}\mbox{\footnotesize #2}}}
\newcommand{\prt}{\partial}
\newcommand{\eps}{\epsilon}

\newcommand{\R}{\mbox{\rule{0.2mm}{2.8mm}\hspace{-1.5mm} R}}
\newcommand{\Z}{Z\hspace{-2mm}Z}

\newcommand{\cd}{{\cal D}}
\newcommand{\cg}{{\cal G}}
\newcommand{\ck}{{\cal K}}
\newcommand{\cw}{{\cal W}}

\newcommand{\vj}{\vec{J}}
\newcommand{\vl}{\vec{\lambda}}
\newcommand{\vz}{\vec{\sigma}}
\newcommand{\vt}{\vec{\tau}}
\newcommand{\vw}{\vec{W}}
\newcommand{\poiss}{\stackrel{\otimes}{,}}

% REVUES POUR BIBLIO

\newcommand{\NP}[1]{Nucl.\ Phys.\ {\bf #1}}
\newcommand{\PL}[1]{Phys.\ Lett.\ {\bf #1}}
\newcommand{\NC}[1]{Nuovo Cimento {\bf #1}}
\newcommand{\CMP}[1]{Comm.\ Math.\ Phys.\ {\bf #1}}
\newcommand{\PR}[1]{Phys.\ Rev.\ {\bf #1}}
\newcommand{\PRL}[1]{Phys.\ Rev.\ Lett.\ {\bf #1}}
\newcommand{\MPL}[1]{Mod.\ Phys.\ Lett.\ {\bf #1}}
\newcommand{\BLMS}[1]{Bull.\ London Math.\ Soc.\ {\bf #1}}
\newcommand{\IJMP}[1]{Int.\ Jour.\ of\ Mod.\ Phys.\ {\bf #1}}
\newcommand{\JMP}[1]{Jour.\ of\ Math.\ Phys.\ {\bf #1}}
\newcommand{\LMP}[1]{Lett.\ in\ Math.\ Phys.\ {\bf #1}}

%\begin{document}
\renewcommand{\thefootnote}{\fnsymbol{footnote}}

\newpage
\setcounter{page}{0}
\pagestyle{empty}

\vs{30}

\begin{center}

{\LARGE {\bf On Coset Reductions of KP Hierarchies}}\\[1cm]

\vs{10}

\vs{10}

{\large {F. Toppan}}\\
\quad \\
{\em Laboratoire de Physique Th\'eorique }\\
{\small E}N{\large S}{\Large L}{\large A}P{\small P}\\
{\em ENS Lyon, 46 all\'ee d'Italie,} \\
{\em F-69364 Lyon Cedex 07, France.}

\end{center}
\vs{20}

\centerline{ {\em  Talk given at Alushta (Crimea), June 1994}}

\vs{20}

\centerline{ {\bf Abstract}}

\indent
In this talk the class of multi-fields reductions of the KP and
super-KP hierarchies (leading to non-purely differential Lax
operators)
is revisited from the point of view of coset construction.
This means in particular that all the hamiltonian
densities of the infinite tower belong to a coset algebra of a given
Poisson brackets structure.
\newpage
\pagestyle{plain}
\renewcommand{\thefootnote}{\arabic{footnote}}
\setcounter{footnote}{0}

\sect{Introduction}

\indent

The hierarchy of integrable equations leading to a solitonic
behaviour has been a widely studied subject since the fundamental
work of Gardner, Green, Kruskal and Miura \cite{GGKM} concerning the
KdV equation. In more recent time it has
particularly deserved physicists' attention expecially in connection
with matrix models, which are a sort of effective description for
two-dimensional gravity.
In such a context the partition functions of matrix models satisfy
the
so-called Virasoro-$\cw$ constraints and can be expressed in terms of
the $\tau$-functions of the hierarchies of classical integrable
equations (for a review on that, see e.g. \cite{mar} and references
therein).\par
Looking at a classification of all possible hierarchies is therefore
a very attracting problem for both mathematical and physical
reasons.
Working in the formalism of pseudo-differential operators (PDO) such
a problem can be formalized as follows: determining
all possible algebraic constraints, consistent with the KP flows, on
the infinite fields entering the KP operator (reduction procedure).
Apart from the well-known solutions of Drinfeld-Sokolov type
\cite{DS}, which can be expressed in terms of purely differential Lax
operators, in the literature other solutions, called
multi-fields KP reductions, have been obtained
\cite{{ara},{bonora},{bonora2}}. Their basic features can be stated
as follows: in the dispersionless limit they give rise
to a Lax operator fractional in the momentum $p$. Moreover, the
algebra of Virasoro-$\cw$ constraints turns out to be a $\cw_\infty$
algebra.\par
Inspired by the works \cite{yuwu} (see also \cite{bakas}), we have
shown
\cite{{toppan},{toppan2}} how such reductions can be derived via a
coset construction, involving a factorization of a Kac-Moody
subalgebra out of a given Kac-Moody
or polynomial $\cw$ algebra. In our framework we have immediately at
disposal a Poisson-brackets structure providing a (multi)hamiltonian
dynamics. Furthermore, the non-linear $\cw_\infty$ algebra can be
compactely interpreted as
a finite rational $\cw$ algebra.\par
In this talk the simplest example, the ${\textstyle{\hat
{sl(2)}}\over {\hat {U(1)}}}$ coset leading to the
Non-Linear-Schr\"{o}dinger equation will be reviewed. The connection
with the modified hierarchy arising from the Wakimoto
representation of ${\textstyle{\hat{sl(2)}}}$ is pointed out.
The formal derivation of our results from the AKS framework to
integrable hierarchies will be given.

\sect{NLS as Coset.}

\indent

 The KP hierarchy (we follow
\cite{dickey}
and the conventions there introduced)
is defined through the pseudodifferential Lax operator
\bea
L &=& \partial + \sum_{i=0}^{\infty} U_i\partial^{-i}
\label{kp}
\ena
where the $U_i$ are an infinite set of fields depending on the
spatial coordinate $x$ and the time parameters $t_k$. Let us denote
as ${L^k}_+$ the purely differential part of the $k$-th power of the
$L$ operator; an infinite set of differential equations, or flows,
for the fields $U_i$ is introduced via the equations
\bea
{\prt L\over \prt t_k}& = &[ {L^k}_+,L]
\label{flows}
\ena
The quantities
\bea
F_k  &=&  <L^k>
\label{first}
\ena
are first integrals of motion for the flows (\ref{flows}). Here the
symbol
$<A>$ denotes the integral of the residue ($<A>=\int dw a_{-1} (w)$)
for the
generic pseudodifferential operator
$A = ...+ a_{-1} \prt^{-1} +... $.\par
 The
reduction procedure
of the KP hierarchy consists in introducing algebraic constraints on
such fields, so that only a finite number of them would be
independent.
Such constraints must be compatible with the flows (\ref{flows}).
Moreover, in order to have hamiltonian dynamics, the constrained
equations should be derived in terms of Poisson bracket structures
and the first integrals of motion should be all in involution with
respect to such P.B. structures.
As
a final result one gets a hierarchy of integrable differential
equations involving
a finite number of fields only.\par
The canonical way to perform a reduction consists in imposing the
constraint
\bea
L^n ={L^n}_+
\label{kdvred}
\ena
which tells that the $n$-th power of $L$ is a purely differential
operator, for a given positive integer $n=2,3,...$ . Such reductions
lead to generalized KdV hierarchies. These hierarchies
are the ones originally described by Drinfeld-Sokolov \cite{DS}.
The set of such reductions given by the constraint (\ref{kdvred})
does not
exhaust the set of all possible reductions compatible with the flows
of KP. It is indeed easily checked that a Lax operator like
\bea
L &=& \partial + V_-\partial^{-1}V_+
\ena
gives rise to consistent KP flows for the fields $V_\pm$.
Let us point out here that the hamiltonian dynamics for such operator
has a nice
interpretation in terms of coset algebra when we realize the
convenience of
replacing the ordinary derivative with a covariant derivative ${\cal
D}$.
Such modification does not affect the KP flows since covariant
derivatives
act on covariant fields following the same rule as ordinary
derivatives.\par
It is remarkable that the above dynamics can be recovered, from a
hamiltonian point of view, once introduced the $sl(2)$ Kac-Moody
algebra as Poisson bracket structure: let us assume
\bea
\{J_+(z),J_-(w)\} &=&  \partial_w\delta(z-w) - 2 J_0 (w) \delta(z-w)
\equiv
\cd (w)\delta(z-w) \nonumber \\
 \{J_0(z), J_\pm (w)\} &=& \pm J_\pm (w) \delta (z-w)
 \nonumber \\
 \{J_0 (z),J_0(w)\} &=& {-\textstyle{1\over
 2}}\partial_w\delta(z-w)\nonumber\\
\{J_\pm(z),J_\pm(w)\} &=& 0
\label{kmalg}
\ena
the covariant derivative $\cal D$ is defined acting on covariant
fields $\Phi_q$ of definite charge $q$ as
\bea
{\cal D} &=& (\partial +2q J_0)\Phi_q
\ena
The property of covariance for the field $\Phi_q$ being defined
through the relation
\bea
\{ J_0 (z), \Phi_q (w) \} &=& q \Phi_q(w)\delta (z-w)
\ena
As its name suggests, the covariant derivative maps covariant fields
of charge $q$ into new covariant fields having the same charge.
In particular $J_\pm $ are covariant fields with respect to $J_0$ and
have charge $\pm 1$ respectively, so that
\bea
{\cal D} J_\pm&= & \partial J_\pm \pm 2 J_0 \cdot J_\pm
\ena
The above introduced reduced Lax operator can be recasted now in the
following form
\bea
L &=& {\cal D} + J_- {\cal D}^{-1} J_+
\equiv \partial +J_-\cdot {\cal
D}^{-1}J_+\label{nls}
\ena
(the last equality when $L$ acts on scalars).\par
The composite bilinear invariant fields $V_n = J_- \cdot {\cal D}^n
J_+$ are a linearly, but not algebraically independent, basis of
generators
for the commutant, the set of fields in the ${\hat{sl(2)}}$
enveloping algebra
having vanishing Poisson brackets with respect to the ${\hat{U(1)}}$
subalgebra:
\bea
\{ J_0 (z), V_n (w) \} &=& 0
\label{comm}
\ena
It is easily checked that the second flow of the reduced KP operator
(\ref{nls})
can be reproduced by using the (\ref{kmalg}) Poisson brackets, and
taking the
second integral of motion as hamiltonian. Moreover all the
hamiltonian densities of the tower automatically belong to the
commutant and are functions of the $V_n$ fields and their
derivatives. So, $H_1 = \int V_0$, $H_2=\int V_1$ and so on.
\par
The above derived hierarchy is the NLS-one, since from the second
flow we obtain the two components NLS equations:
\bea
{\dot {J_\pm}}&= &  \{ J_\pm , H_2\}_1=\pm {\cal D}^2 J_\pm \pm 2
(J_+ J_-)J_\pm
\label{nls2}\ena
together with ${\dot {J_0}} =0$. Setting $J_0\equiv 0$ we obtain
precisely th standard form of the two components NLS.\par
In our framework we do not need to perform any Dirac reduction, for
that reason in particular the connection with the m-NLS hierarchy is
very transparent.
As in the KdV-type case, the modified hierarchies are obtained once
introduced a free-field representation. In our case the free-field
representation is provided by the Wakimoto fields \cite{waki}, namely
the weight $1$ field ${\nu}$ and the bosonic $\beta-\gamma$ system
of weight $(1,0)$, satisfying the free field algebra
\bea
\{\beta (z) , \gamma (w) \} &=& -\{ \gamma (z) , \beta (w) \} =
\delta (z-w)\nonumber\\
\{\nu (z), \nu (w) \} &=& \partial_w \delta (z-w)
\ena
(any other Poisson bracket is vanishing).\par
The $sl(2)-{\cal KM}$ algebra given in (\ref{kmalg}) is reproduced
through
the identifications
\bea
J_+&=&\beta\nonumber\\
J_0 &=& -\beta \gamma + {i\over {\sqrt 2}}\nu \nonumber\\
J_- &=& \beta \gamma^2 -i{\sqrt 2} \gamma \nu +\partial\nu
\ena
Expressing the $H_2 $ hamiltonian in terms of the Wakimoto fields,
one can derive the m-NLS equation for the $\beta,\gamma,\nu$ fields.
This is nothing else that the generalization of the Miura
transformation to this new class
of reductions.\par
The above scheme does not apply only in that particularly simple
case, but is
completely general: the next simplest example of consistent KP
reduction is a $3$-fields reduction, given by the Lax operator
\bea
{\tilde L}& =& {\cal D}^2 + T + W_-\cdot {\cal D}^{-1} W_+
\label{op}
\ena
where $T, W_\pm$, together with a spin $1$ field $J(z)$ entering the
covariant derivative, form a closed algebra, the
non-linear Polyakov-Bershadski
$\cw$ algebra \cite{polya},
which provides a Poisson brackets structure and the hamiltonian
dynamics for the
corresponding flows.
Let me just mention that in the limit of dispersionless
Lax equation, the reduced operators (\ref{nls})
and (\ref{op}) are respectively given by
\bea
L&\rightarrow& p + {\lambda \over p+ \alpha} \nonumber\\
{\tilde L}& \rightarrow & p^2 + t +{\lambda \over p+\alpha}
\ena
($\alpha ,\lambda$ and $ t $ constants),
namely by fractions in the momentum $p$ and not polynomials like
standard KdV-type reductions. This is a common feature of all coset
reductions.\par
Let us point out that the above scheme can be introduced even in the
supersymmetric case, when considering reductions of the Manin-Radul
super-KP operator\cite{manin}. For further details see
\cite{toppan2}.

\section{The AKS approach.}

\indent

The above derived results can be formally proved in the
general case
once realized that they can be formulated in terms of the AKS
approach to the hierarchies\footnote{In this section I will talk
about a work in preparation, done in collaboration with L. Feher}.
In this scheme one starts with a matrix-type Lax operator of the kind
\bea
{\cal L }&=& \partial_x +J(x) + \lambda K
\ena
where $J(x)$ are currents valued in some finite Lie algebra ${\cal
G}$, $\lambda$ is a
spectral parameter and K a constant element in ${\cal G}$.
The Kac-Moody current algebra ${\hat {\cal G}}$ is one of the Poisson
brackets structure for the above system (the one we are interested
in).
The generalized Drinfel-Sokolov hierarchies are obtained (see
\cite{genkdv}) by assuming $K$ being a regular element for the loop
algebra ${\tilde{\cal G}}$,
where the loop parameter is the spectral parameter. The regularity
condition
reads as follows
\bea
{\tilde{\cal G}} &=& Ker K \oplus Im K
\ena
where the action is the adjoint one.\par
Under the above assumption it is possible to introduce a grading
$deg$ on the elements of the loop algebra, defined in such a way that
$deg (\lambda K) =1$.
Different gradings induce different hierarchies; the principal
grading, which associates
the grade one to the simple positive roots of the algebra, provides
the standard DS hierarchies. Another interesting class of gradings is
provided by the homogeneous one, introduced through $deg \equiv
{\textstyle {\lambda d\over d\lambda }}$. The NLS hierarchy is
obtained by taking the homogeneous
grading w.r.t. the ${\cal G} \equiv sl(2)$ algebra having $H$ as
Cartan generator and $E_\pm$ as roots. In this case $K=H$ is a
regular element.\par
The main property of the above operators can be stated as follows:
there exists an adjoint transformation
${\cal L} \mapsto {\cal L}_\alpha = adj (\alpha ) {\cal L}$,
\bea
{\cal L}_{\alpha} &=& {\cal L} +[\alpha ,{\cal L}] +{\textstyle
{1\over 2}}[\alpha , [\alpha,{\cal L}],] +...
\ena
which preserves both the Poisson brackets and the monodromy
invariants.
Under the condition of regularity for $K$, it is possible to
determine uniquely,
with an iterative procedure,
the local fields $\alpha (x)\in Im K$, $\alpha (x) $ being expanded
in the components with negative grading only, such that the
transformed operator
${\cal L}_\alpha $ is diagonal
\bea
{\cal L}_\alpha &=& \partial_x + R(x) + \lambda K
\ena
where $R(x)\in Ker K $ is expanded over the non-positive grading
components only
and can be iteratively computed. The diagonal character of the
transformed operator makes possible to compute the monodromy
invariants. The different components of $R(x)$ provide the tower of
hamiltonian densities.
In the NLS case we have, e.g.,
\bea
\alpha (x) = \lambda^{-1}\alpha_1 +\lambda^{-2}\alpha_2 +...
\nonumber
\ena
with
\bea
\alpha_j &=& \alpha_{j,+}E_+ + \alpha_{j,-}E_-\nonumber
\ena
and
\bea
R(x) &=& R_0 H + \lambda^{-1}R_{1} H +\lambda^{-2} R_{2} H +...
\ena
It is easy to verify that, at the lowest orders
\bea
{\alpha_{1,\pm}}&=& \pm J_\pm\nonumber\\
\alpha_{2,\pm} &=& -{\cal D} J_\pm \nonumber\\
\alpha_{3,\pm} &=& \pm {\cal D}^2J_\pm \mp {\textstyle{4\over 3}
(J_+J_-)J_\pm
\ena
and
\bea
R_0 &=& J_0\nonumber\\
R_1 &=& 2 J_+J_-\nonumber\\
R_2 &=& J_+{\cal D} J_- - J_- {\cal D} J_+
\ena
Therefore $R_2$ is the hamiltonian density giving rise to the NLS
equation w.r.t. the Kac-Moody Poisson brackets.
It needs just two pages of computation to prove by induction that
the fields ${\alpha_{j,\pm}}$ are covariant with charges $\pm1$
respectively and that all the $R_j$'s apart from $R_0$ belong to the
coset.\par
The above considerations can be clearly repeated for more complicated
hierarchies to show that coset structures arise naturally.
\par
There is one point we have left apart, namely how
to connect scalar Lax operators $L$ of the type considered in the KP
framework,
with the matrix ones (denoted by ${\cal L}$). \\
Let us take for simplicity the $sl(2)$ case. In the fundamental
representation
we can consider the matrix equation
\bea
{\cal L}\Psi = 0 ={\cal D} \Psi =0 &\equiv& \nonumber\\
\left(\partial + \left( \begin{array}{cc}
J_0 & J_+ \\
J_- & -J_0 \end{array} \right)\right)
\left( \begin{array}{c} \Psi_+\\ \Psi_- \end{array}\right) &=&0
\ena
If we solve it for, let's say, the $\Psi_-$ component and we allow
inverting the derivative operator, then we can plug the result into
the equation for the
$\Psi_+$ component;
we get
\bea
({\cal D} +J_-{\cal D}^{-1}J_+ ) \Psi_+ &=& 0
\ena
namely we are led with the scalar Lax operator previously
considered.\par
Notice that, instead of starting with the fundamental (spin
${\textstyle{1\over 2}}$) representation of
$sl(2)$, we can choose any other representation, for instance the
triplet (acting on the vector ($\Psi_{+1},\Psi_0,\Psi_{-1}$).
Proceding as before we are led in this case to a different equation:
in terms of the $\Psi_0$ component it reads as
\bea
({\cal D} +J_-{\cal D}^{-1}J_+
+J_+{\cal D}^{-1}J_- ) \Psi_0 &=& 0
\ena
The scalar Lax operator we obtain is put in a different, but
equivalent form,
with respect to the previous one. It is basically the symmetrized
form
under the exchange $J_-\leftrightarrow J_+$. This is a standard
feature
of the derivation of scalar Lax operators out of matrix ones. There
is no canonical way of doing it, instead a class of different but
equivalent
such operators can be found.

{}~\\~\\

\noindent
{\large{\bf Acknowledgements}}
{}~\\~\\
I should acknowledge useful discussions had with I. Krichever and L.
Feher.
{}~\\
{}~\\


\begin{thebibliography}{99}

\bibitem{GGKM} C.S. Gardner, J.M. Green, M.D. Kruskal and R.M. Miura,
Phys. Rev. Lett. 19 (1967), 1095.
\bibitem{mar} A. Marshakov, Int. Jou. Mod. Phys. A, Vol.8, n. 22
(1993), 3831.
\bibitem{DS} V. Drinfeld and V. Sokolov, Jou. Sov. Math 30 (1984),
1975.
\bibitem{ara} H. Aratyn, L.A. Ferreira, J.F. Gomes and A.H.
Zimerman,
Nucl. Phys. {B 402} (1993), 85; Preprint UIC-HEP-TH-93-05, hep-th
9304152.  H. Aratyn E. Nissimov and S. Pacheva, Phys. Lett. B 314
(1993) 41;
Preprint UICHEP-TH-94-2, hep-th 9401058.
\bibitem{bonora} L. Bonora and C.S. Xiong, Phys. Lett. B 285 (1992),
191;
Phys. Lett. {B 317}
(1993), 329.
\bibitem{bonora2} L. Bonora and C.S. Xiong, Preprint SISSA-171/93/EP,
BONN-HE-46/93, hep-th 9311070.
\bibitem{yuwu} F.~Yu and Y.S.~Wu, Phys. Rev. Lett. {68} (1992),
2996; Nucl. Phys. {B 373} (1992), 713.
\bibitem{bakas} I. Bakas and E. Kiritsis, Int. J. Mod. Phys. A7
(1992), 55;  J. Schiff, Preprint IASSNS-HEP-92-57, hep-th
9210029.
\bibitem{toppan} F. Toppan,  Phys. Lett. B 327 (1994), 249.
\bibitem{toppan2} F. Toppan, Preprint Enslapp L-467/94, hep-th
940595,
to appear in Int. Jou. Mod. Phys. A.
\bibitem{dickey} L.A. Dickey, Soliton Equations and Hamilton Systems,
Adv. Series in Math. Phys. Vol. 12, World Sc. 1991.
\bibitem{waki} M. Wakimoto, Comm. Math. Phys. 104 (1986), 605.
\bibitem{polya} A. Polyakov, Int. J. Mod. Phys. {A 5} (1990),
833; M.~Bershadsky, Comm. Math. Phys. 139 (1991) 71.
\bibitem{manin} Yu.I. Manin and A.O. Radul, Comm. Math. Phys. 98
(1985), 65.
\bibitem{genkdv} I. Bakas and D.A. Depireux, Int. Jou. Mod. Phys. A7
(1992), 1767;
M.F. de Groot, T.J. Hollowood and J.L. Miramontes,
Comm. Math. Phys. {145} (1992), 57; L. Feher, J. Harnad and I.
Marshall, Comm. Math. Phys. {154} (1993), 181; L. Feher, preprint
LANL -hep-th 9211094.
\end{thebibliography}
\end{document}